\documentclass[twocolumn]{aastex6}

\usepackage{amsmath}
\usepackage{natbib}
\usepackage{threeparttable}

\usepackage{cleveref}
\crefname{figure}{Figure}{Figures}

\newcommand{\be}{\begin{equation}}
\newcommand{\ee}{\end{equation}}

\newcommand{\ba}{\begin{eqnarray}}
\newcommand{\ea}{\end{eqnarray}}

\newcommand{\Wi}{\ensuremath{W_i}}

\newcommand{\Wij}{\ensuremath{W_{ij}}}

\newcommand{\Wrij}{\ensuremath{\mathcal{W}_{ij}^R}}
\newcommand{\Wrji}{\ensuremath{\mathcal{W}_{ji}^R}}

\newcommand{\vv}{\ensuremath{\boldsymbol{v}}}
\newcommand{\vx}{\ensuremath{\boldsymbol{x}}}
\newcommand{\half}{\ensuremath{\frac{1}{2}}}
\newcommand{\sumj}{\ensuremath{\displaystyle\sum_j}}

\newcommand{\etacrit}{\ensuremath{\eta_{\text{crit}}}}
\newcommand{\etafold}{\ensuremath{\eta_{\text{fold}}}}

\newcommand{\Pilin}{\ensuremath{\Pi_{\text{lin}}}}
\newcommand{\Piquad}{\ensuremath{\Pi_{\text{quad}}}}

\newcommand{\eg}{\textit{e.g.\ }}

\DeclareMathOperator{\Tr}{Tr}
\DeclareMathOperator{\sign}{sign}

\begin{document}
\title{Examining the Accuracy of Astrophysical Disk Simulations With a Generalized Hydrodynamical Test Problem}

\author{Cody Raskin and J.~Michael Owen}
\affil{Lawrence Livermore National Laboratory, P.O. Box 808, L-038, Livermore, CA 94550}
\email{raskin1@llnl.gov}
\email{mikeowen@llnl.gov}

\begin{abstract}
We discuss a generalization of the classic Keplerian disk test problem allowing for both pressure and rotational support, as a method of testing astrophysical codes incorporating both gravitation and hydrodynamics.
% for use in studying astrophysical disks.
We argue for the inclusion of pressure in rotating disk simulations on the grounds that realistic, astrophysical disks exhibit non-negligible pressure support.
We then apply this test problem to examine the performance of various smoothed particle hydrodynamics (SPH) methods incorporating a number of improvements proposed over the years to help SPH better address problems noted in modeling the classical gravitation only Keplerian disk.
We also apply this test to a newly developed extension of SPH based on reproducing kernels called CRKSPH.
Counterintuitively, we find that pressure support worsens the performance of traditional SPH on this problem, causing unphysical collapse away from the steady-state disk solution even more rapidly than the purely gravitational problem, whereas CRKSPH greatly reduces this error.
\end{abstract}

\section{Introduction}
%\listoftodos

Rotating disks are one of the most ubiquitous astrophysical phenomena in the universe. 
Besides being the primary component of spiral galaxies, they appear as quasi-stable, thermalized disks following compact object mergers and collisions \citep{rosswog2009wd,rosswog2010,vanKerkwijk2010,raskin2010,raskin2012,raskin2014}, as nucleating disks during stellar and planetary formation\citep{cassen1981,lissauer1987,lubow1999,boss2001,imaeda2002,saitoh2004,saitoh2008,krumholz2009,wada2011}, and in the form of accretion disks around a variety of objects, such as black holes \citep{chakrabarti1993,murray1996,fryer1999,rosswog2008,rosswog2009bh,rosswog2010bh}. 
All of these scenarios are topics of great interest in the astrophysical community, and as a consequence, a considerable amount of computational effort has been expended modeling these objects using a variety of methods \citep{flebbe1994,owen1998,gadget,boley2007,cullen2010,hopkins2015,beck2016}. 
One important class of such computational methods are the various forms of Smoothed Particle Hydrodynamics \citep[SPH,][]{gingold1977,lucy1977}.

It is therefore no surprise that many papers describing astrophysical hydrodynamics methods -- SPH included -- employ a simplified, rotating astrophysical disk as a test of the method \citep{owen1998,owen2004,lodato2004,cullen2010,hopkins2015,beck2016}.
The most common example of this sort of rotating astrophysical test is the classic Keplerian disk \citep{masuda1997,cartwright2010,binney2011,hosono2016}.
However, the Keplerian disk scenario is a purely gravitational problem, such that this is more a test of how well the gravitational portion of the algorithm can properly model the orbital motion of the fluid, with no role for hydrodynamics other than that it not interfere with the gravitational problem.
While such a test is an important limit to examine, it does not represent a rigorous test of astrophysical hydrodynamics, and neglects to test an important physical regime.
Pressure plays a role in virtually every astrophysical disk to various degrees (either in their formation or providing some degree of support/equilibrium), and the applicability of any hydrodynamic method for the study of astrophysical disks depends greatly on the degree to which 
the method is able to properly simulate these disks in the presence of non-negligible pressure. 

Including pressure in a rotating disk simulation can expose potential weaknesses of a method, especially in the presence of shearing flows which are typical of a rotationally supported astrophysical disk.
The improper activation of the SPH artificial viscosity in the presence of shearing flows is one such well-known weakness that can degrade an SPH simulation of a rotating disk.
Because the purely gravitational Keplerian disk is the most common test problem used to investigate how well SPH models can represent astrophysical disks, 
a considerable amount of effort in the literature has been expended to ensure that SPH methods act more like $n$-body codes in these kinds of tests \citep{balsara1995,cullen2010}.
This focus unfairly neglects the full hydrodynamic role of pressure, and tells us nothing of how applicable SPH might (or might not) be to truly model the formation and evolution of such disks.
SPH is a full-featured hydrodynamics method, and as hydrodynamics plays a non-negligible role in nearly all physical scenarios involving a gravitating disk, we would like a broader class of test problems -- a treatment where both the kinematics and hydrodynamics are modeled accurately and robustly.

In this paper, we describe a simple gravitationally bound, rotating disk test case that permits an arbitrary fraction of pressure vs.~rotational support, and we demonstrate how a variety of formulations of SPH fare on this problem, including a new variation based on a reproducing kernel formulation \citep[CRKSPH,][]{frontiere2016}. 
In \S\ref{sec:Qforms}, we briefly review the role of artificial viscosity in SPH and the various modifications that exist to ameliorate its effects on shear flows. 
We also provide in \S\ref{sec:Qforms} a brief review of the framework upon which CRKSPH is built. In \S\ref{sec:disktest}, we outline a generalization of the Keplerian disk problem we use to create a general class of pressure-supported, rotating disks. 
In \S\ref{sec:results} we discuss the performance of the various competing artificial viscosity forms against CRKSPH, and in \S\ref{sec:conclusion}, we give our conclusions.

\section{Artificial Viscosity in SPH}
\label{sec:Qforms}
Some form of artificial viscosity is a required component for SPH in order to properly capture shock physics -- for our purposes we sweep Riemann-solver based SPH into this same category, as Riemann-solvers and artificial viscosities are closely related.
Put another way, the SPH discretization scheme requires that the momentum field be differentiable everywhere, and that is only true in the presence of shocks with the inclusion of some kind of artificial viscosity (see \cite{price2005} for a review) that is activated by convergent velocity flows. 
These dissipative terms (in the form of either an artificial viscosity or Riemann-solver) act to relax discontinuities to the resolution scale of the SPH simulation. 
The most commonly used artificial viscosity is the version developed by \cite{monaghan1983}, which modifies the standard inviscid momentum conserving equation with an artificial viscosity $\Pi$, such 
that for the $i$th particle,
\be
\label{eq:momentum}
\frac{d\vv_i}{dt} = -\sumj m_j\left(\frac{P_i}{\rho_i^2} + \frac{P_j}{\rho_j^2} + \Pi_{ij}\right){\nabla}_iW_{ij},
\ee
where $m$, $P$, and $\rho$ have the usual meanings for mass, pressure, and density, respectively, and $W$ is the SPH smoothing kernel function. 
In this case $\Pi_{ij} = (\Pi_i + \Pi_j)/2$, and $\Pi_{i}$ takes the form
\begin{align}
  \label{eq:MGPi}
  \Pi_i &= \rho_i^{-1} \left(-C_l c_i \mu_i + C_q \mu_i^2 \right) \\
  \label{eq:MGmu}
  \mu_i &= \min\left(0,\frac{h_i\vv_{ij}\cdot\vx_{ij}}{|\vx_{ij}|^2 + \epsilon h_i^2}\right).
\end{align}
Here, $C_l$ and $C_q$ are tunable parameters for the linear and quadratic functions of the velocity difference, respectively, $c_i$ is the sound speed,  $h_i$ is the SPH smoothing length, 
$\vv_{ij} \equiv \vv_i - \vv_j$, $\vx_{ij} \equiv \vx_i - \vx_j$, and $\epsilon$ is a small number to avoid division by zero. The
commensurate specific thermal energy equation consistent with \cref{eq:momentum} is 
\be
\label{eq:energy}
\frac{du_i}{dt} = \sumj m_j\left(\frac{P_i}{\rho_i^2}+\frac{1}{2}\Pi_{ij}\right)\vv_{ij}\cdot\nabla_i W_{ij}.
\ee
We will henceforth refer to this type of viscosity as M\&G.

It is worth noting the energy evolution may be handled by \cref{eq:energy}, a total energy formulation, the compatible energy update of \citep{owen2014}, replaced by an entropy relation, or various other ways described in the literature.
In this paper we use the compatible energy update of \citep{owen2014}, but this detail is not critical to the results we find.
There are also various other ways to symmetrize the conservation equations other than the choices of \crefrange{eq:momentum}{eq:energy} that we do not cover here \citep[see \eg][]{gadget,fryer2006,owen2014,frontiere2016}.

The major drawback of the M\&G viscosity is that it results in excess entropy production, as it is active for any compressional flow (or more precisely any interacting pair of SPH points that are approaching one and other), as opposed to just in the presence of a shock. 
Over the years, there have been a number of efforts to modify this simple scheme so as to limit its deleterious effects outside of shocks. 
We briefly review these efforts below.

\subsection{Morris \& Monaghan}

\cite{morris1997} suggested a method to limit the artificial viscosity  whereby the magnitude of $\Pi_i$ is governed by a new parameter $\alpha_i$ with its own evolution equation;
\be
\frac{d\alpha_i}{dt} = \frac{\alpha_i-\alpha^*}{\tau_i} + S_i.
\ee
This equation acts like a damping function where $\alpha$ decays to a constant ($\alpha^*\sim0.1$) over an $e$-folding time $\tau$. $S$ is a source function that detects the 
presence of a shock, typically by $S_i=\max\{-{\nabla}\cdot\vv_{i},0\}$, and the characteristic decay timescale is typically set to $\tau_i\approx h_i/(0.2c_i)$. Outside of shocks, 
this damping function drives the magnitude of the artificial viscosity parameter toward $\alpha_i \to \alpha^*$, which is chosen to be a low value that balances the utility of artificial viscosity in limiting
sub-resolution scale noise and spurious particle inter-penetration against the non-desirous entropy generation outside of shocks. We will henceforth refer to this viscosity limiting scheme as M\&M
viscosity.

One serious deficiency in this approach is that by governing the limiter with an evolution equation that reacts to a source term, there is a built-in delay between the appearance of a shock and the
activation of the full-strength artificial viscosity. This results in oscillatory Gibbs phenomena in the post-shock region \citep{gibbs1898}. Though this is not a concern for the specific test problem presented 
in this paper -- that of a frictionless rotating disk -- this behavior has nevertheless argued against the widespread adoption of the M\&M limiter. Moreover, as this merely limits the M\&G viscosity
magnitude to $\alpha^*$ even outside of shocks, M\&M viscosity is still overly viscous in shearing flows.

\subsection{Balsara Switch}

Preceding \cite{morris1997}, \cite{balsara1995} suggested leveraging some of the information from the full velocity gradient to develop a limiter on the pair-wise M\&G type viscosity.
This limiter takes the form
\be
\label{eq:balsara}
\alpha_i = \frac{|\nabla\cdot\vv_i|}{|\nabla\cdot\vv_i|+||\nabla\times\vv_i||+\epsilon c_i/h_i}.
\ee

Typically, this factor is multiplied on $\Pi_{ij}$ as the average of $\alpha_i$ and $\alpha_j$. When the curl of the velocity field is high, but the divergence is low, this factor reduces the
total artificial viscosity strength, limiting its effect in vortical flows.
However, this multiplier is really only correct in pure shear flows (where it turns off the viscosity) or purely compressing flows (where it leaves the viscosity full strength).
It does not do anything to correct the activation of the viscosity in compressing but non-shocking flows, nor is it correct in the presence of combined shocks and shears.
There is also the concern that it can be degraded by an inaccurate measurement of the velocity gradient, as noted in \cite{cartwright2009}.
In our SPH implementation however, we employ a more accurate, linearly corrected velocity gradient (discussed below in \S\ref{sec:gradv}), so this particular concern is alleviated.

One additional concern is that the Balsara switch may suppress artificial viscosity in the case of pathological motion of a single particle in the presence of 
vortical flow. In other words, in a strong shear scenario where any single particle might develop spuriously high velocity, the kernel-smoothed divergence of the velocity field could hide this motion
resulting in a small value for $\alpha$, thus suppressing viscosity when we would like to dissipate such a highly local, pair-wise extreme velocity jump.

\subsection{Cullen \& Dehnen Switch}

\cite{cullen2010} developed their own version of a limiting switch that is a combination of the previous two approaches (henceforth, we will refer to this approach as simply the C\&D viscosity). 
Their source term equivalent to $S_i$ in the M\&M viscosity takes the form
\be
\label{eq:cullen}
S_i = \xi_i \max\{-\dot{\nabla}\cdot\vv_i,0\},
\ee
where $\xi_i$ is a compression detection limiter function, and $\dot{\nabla} = d\nabla/dt$. The limiter function has the form
\ba
\label{eq:cullenxi}
\xi_i &=& \frac{\left|2(1-R_i)^4\nabla\cdot\vv_i\right|^2}{\left|2(1-R_i)^4\nabla\cdot\vv_i\right|^2 + \Tr(\mathbf{S}_i\cdot\mathbf{S}_i^\top)},\\
\label{eq:cullenri}
R_i &\equiv& \frac{1}{\hat{\rho}_i} \sumj\sign(\nabla\cdot\vv_j)m_j W(|\vx_{ij}|,h_i),
\ea
with $\mathbf{S_i}$ representing the shear component of the velocity gradient matrix $\mathbf{V} \equiv \nabla \otimes \vv$, namely the traceless symmetric component 
$\mathbf{S}\equiv (\mathbf{V}+\mathbf{V}^\top)/2 - \nu^{-1}(\nabla\cdot\vv)\mathbf{I}$, where $\nu$ is the number of spatial dimensions. 

Rather than integrating a differential equation and thus ramping up in the presence of a shock as is done in the M\&M viscosity, the C\&D viscosity switch sets $\alpha_i$ to 
\be
\label{eq:alphaloc}
\alpha_i^\prime = \alpha_{\max} \frac{h_i^2 S_i}{v^2_{{\rm sig},i}+h_i^2 S_i}
\ee
whenever this quantity exceeds the existing value for $\alpha_i$. The signal velocity is given by
\be
\label{eq:vsig}
v_{{\rm sig},i} = \displaystyle{\max_{|\vx_{ij}|\leq h_i}}\{\bar{c}_{ij}-\min\{0,\vv_{ij}\cdot\hat{\vx}_{ij}\}\}.
\ee
If $\alpha_i^\prime<\alpha_i$ at any time step, $\alpha_i$ decays according to $\dot{\alpha_i} = (\alpha_i^\prime - \alpha_i)/\tau_i$, where $\tau_i = h_i/0.1v_{{\rm sig},i}$.

Formulated this way, with an instantaneous maximum and a steady decay after a shock, the C\&D viscosity features reduced Gibbs phenomena over the M\&M formulation. Moreover,
with the addition of the limiter function $\xi$, which depends on the shear component of the velocity, the C\&D viscosity is better able to reduce or eliminate viscosity in purely shearing flows. 

\subsection{Linearly Corrected Velocity Gradient}
\label{sec:gradv}
For this paper, in order to improve the accuracy of the velocity gradient to be formally exact for linear velocity fields \citep{Randles1996}, we replace the standard SPH velocity gradient with a linearly corrected gradient of the form
\begin{align}
  \label{eq:SPHDvDx}
  \nabla\cdot\vv_i &= -M_i^{-1} \sum_j m_j \vv_{ij}\cdot\nabla\Wi,\\
  \label{eq:SPHM}
  {M}_i &= -\sum_j m_j \vx_{ij}\cdot\nabla\Wi.
\end{align}
This gradient is used for both the Balsara and C\&D viscosities, improving their accuracy and results over that obtained by using the uncorrected SPH velocity gradient.

\subsection{Reproducing Kernels and the Christensen Viscosity}

Recently, \cite{frontiere2016} demonstrated a method for incorporating reproducing kernels \citep[RPK, see \eg][]{liu1995,liu1998} into SPH while maintaining the conservation properties of
traditional SPH. Briefly, the crux of their method (called CRKSPH) is first the adoption of a reproducing form of the kernel and its gradient,
\begin{align}
  \label{eq:wR}
  \Wrij(\vx_{ij}) \equiv& A_i \left(1+\boldsymbol{B}_i \cdot\vx_{ij}\right)\Wij(\vx_{ij}), \\
  \label{eq:gradwR}
  \nabla\Wrij(\vx_{ij}) =& A_i \left(1+\boldsymbol{B}_i \cdot\vx_{ij}\right)\nabla \Wij(\vx_{ij}) \nonumber \\
  &+ \nabla A_i\left(1+\boldsymbol{B}_i \cdot\vx_{ij}\right)\Wij(\vx_{ij}) \nonumber\\
  &+ A_i\left(\nabla(\boldsymbol{B}_i \cdot\vx_{ij}) + \boldsymbol{B}_i\right)\Wij(\vx_{ij}),
\end{align}
where $A_i$ and $\boldsymbol{B}_i$ are the zeroth and first-order corrective coefficients. These coefficients are derived from the moments of the particle positions, with
\begin{align}
  \label{eq:coefA}
  A_i &= \left[m_0-\left(\boldsymbol{m}_2^{-1}\cdot \boldsymbol{m}_1\right)\cdot \boldsymbol{m}_1\right]^{-1} \\
  \label{eq:coefB}
  \boldsymbol{B}_i &= -\boldsymbol{m}_2^{-1}\cdot \boldsymbol{m}_1,
\end{align}
where the geometric moments are
\begin{align}
  \label{eq:m0}
  m_0 &\equiv \sum_j V_j \Wij \\
  \label{eq:m1}
  \boldsymbol{m}_1 &\equiv \sum_j \vx_{ij} V_j\Wij \\
  \label{eq:m2}
  \boldsymbol{m}_2 &\equiv \sum_j \vx_{ij} \otimes \vx_{ij} V_j\Wij.
\end{align}
In this case, $V_j$ is the volume of the $j$th particle. The corrections described here and in \cite{frontiere2016} are for linear-order corrected kernels, but the same formalism 
can be used for any order correction one desires. The corresponding, conservative evolution equations (in inviscid form) are
\begin{align}
  \label{eq:momeq2}
  m_i \frac{d\vv_i}{dt} &= -\half \sum_j V_i V_j (P_i + P_j) \left( \nabla \Wrij  - \nabla \Wrji \right), \\
  \label{eq:engeq2}
  m_i \frac{du_i}{dt} &= \half \sum_j V_i V_j P_j \vv_{ij} \cdot\left( \nabla \Wrij - \nabla \Wrji \right),
\end{align}
and we refer to \cite{frontiere2016} for their derivations, for the forms of $\nabla A$ and $\nabla \boldsymbol{B}$, and for any other numerical details not covered here.

Employing the RPK corrected velocity gradient
\begin{equation}
  \nabla\cdot\vv_i = -\sum_j V_j \vv_{ij} \cdot\nabla \Wrij,
\end{equation}
\cite{frontiere2016} construct new limiter for $\vv_{ij}$ in \cref{eq:MGmu}, inspired by the limiter methods used in \cite{christensen1990}. This limited value, $\hat\vv_{ij}$, is projected to the midpoint between particles $i$ and $j$ via
\begin{align}
  \label{eq:viLim}
  \hat{\vv}_i &\equiv \vv_i + \half \phi_{ij} \nabla\otimes\vv_i \cdot\vx_{ji}  \\ 
  \label{eq:vjLim}
  \hat{\vv}_j &\equiv \vv_j + \half \phi_{ji} \nabla\otimes\vv_j \cdot\vx_{ij}  \\ 
  \label{eq:vDiff}
  \hat\vv_{ij} &\equiv \hat{\vv}_i - \hat{\vv}_j,
\end{align}
where $\phi_{ij}$ is a van Leer limiter function \citep{vanLeer1974} of the form
\begin{align}
  \label{eq:phi}
  \phi_{ij} &= \max\left[0, \min\left(1, \frac{4 r_{ij}}{(1 + r_{ij})^2}\right)\right]\nonumber\\
   &\times
  \left\{ \begin{array}{l@{\quad}l}
    \exp\left(-\left((\eta_{ij} - \etacrit)/\etafold\right)^2\right), & \eta_{ij} < \etacrit \\
    1, & \eta_{ij} \ge \etacrit \\
  \end{array} \right. \\
  \label{eq:rij}
  r_{ij} &\equiv \frac{\nabla\otimes \vv_i\cdot \vx_{ij}\cdot\vx_{ij}}{\nabla\otimes \vv_j\cdot \vx_{ij}\cdot\vx_{ij}} \\
  \eta_{ij} &\equiv \frac{\sqrt{\vx_{ij}\cdot\vx_{ij}}}{\max\left( h_i, h_j \right)}.
\end{align}

CRKSPH was developed in response to some of the unavoidable errors associated with the traditional SPH method, chief among these being the inaccurate interpolation theory underlying SPH (i.e. the so-called ``E0'' error whereby even a constant field is not in general interpolated correctly), and the overly aggressive artificial viscosity schemes currently available.
Reproducing kernel theory is explicitly designed to interpolate more accurately than the corresponding SPH identities, and the higher-order projection used to construct the CRKSPH artificial viscosity outlined in \crefrange{eq:viLim}{eq:rij} are designed to correct the viscosity limitations of SPH.

\cite{frontiere2016} examine the performance of CRKSPH on a variety of problems, including shocks, shearing flows, vortical motion, instabilities and combinations therein. 
They find that CRKSPH is at least as performant as any of the modern, competing forms of SPH on a number of tests, and in most cases yields the best result.
While they do examine the performance of CRKSPH on a pair of classic vortex tests (the Gresho and Yee vortex tests, \citep{gresho1990,Yee2000}), which have some similarities to a gravitational disk, they do not explicitly test a more astrophysically motivated, Keplerian-type disk with a central, gravitational potential.

\section{A Pressure-Supported Rotating Disk}
\label{sec:disktest}
For the sake of gauging the relative effect of pressure on the numerical stability of a rotating disk, we choose to construct a disk whose pressure can be modulated with a single parameter ($f_p$) across the entirety of the disk, as was done in \cite{owen1998} and \cite{owen2004}.
This constraint fixes the density profile of the disk to be commensurate with the gravitational potential and the chosen equation of state. 
To start, we simply balance the force of gravity against both the pressure force and the centripetal force,

\be
\frac{\nabla P}{\rho} - \frac{v_\theta^2}{r}= \nabla\Phi,
\ee
using a softened gravitation potential of the form
\be
\Phi(r) = -\frac{GM}{\left(r^2+r_s^2\right)^{1/2}}\label{eq:potential}.
\ee
Note that in these relations we neglect any self-gravitation of the disk material: the gravitational potential is entirely due to this central softened source.

If we use a polytropic equation of state, $P(r)= K \rho^\gamma$, where $\gamma=(n+1)/n$, then the density profile reduces to
\be
\rho(r) = \left[\frac{GM(\gamma-1)}{K\gamma (r^2+r_s^2)^{1/2}}\right]^{1/(\gamma-1)}\label{eq:density}.
\ee
We choose $n=2$, $\gamma=3/2$ for simplicity, yielding
\be
K=\frac{GM}{3r_s\rho_0^{1/2}}.
\ee
At this stage, $P$ can be modulated by a scalar parameter ($f_p$) such that a value of 1 results in a fully pressure-supported, static disk. $v_\theta$ then becomes
\be
v_\theta^2 = (1-f_p)\frac{GMr^2}{(r^2+r_s^2)^{3/2}}\label{eq:velocity}.
\ee
Note that in the special case $f_p=r_s=0$, \cref{eq:velocity} reduces to the well known Keplerian circular velocity $v_\theta^2 = GM/r$.
Therefore these relatations represent a generalization of the classical Keplerian disk, admitting an arbitrary degree of pressure support. 

We now have all of the components we need to build pressure-supported, rotating disks. Additionally, formulating our disks this way ensures that adjusting the relative strengths of 
pressure and velocity against gravity does not alter the matter distribution of our disk. Thus, we have effectively decoupled the effects of pressure and viscosity on the evolution of 
our disks from the arrangement of our disk material, and we can more easily isolate how SPH handles shearing flows, and to what degree it becomes overly diffusive. 

\begin{figure}[ht]
\centering
\includegraphics[width=0.40\textwidth]{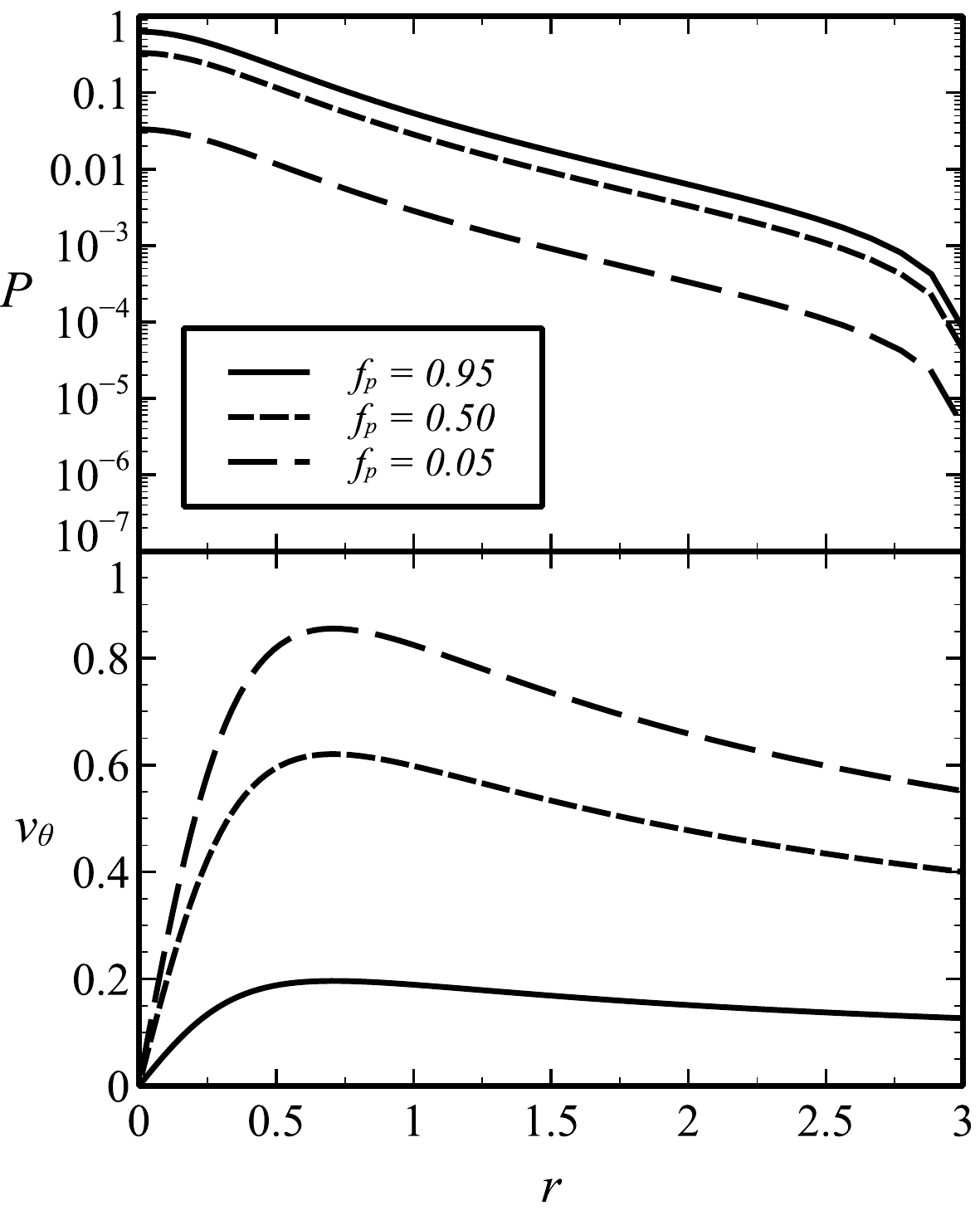}
\caption{Analytical profiles for pressure and angular velocity for different values of $f_p$.}
\label{fig:profiles}
\end{figure}

For all of our tests, we set $\rho_0=GM=1.0$, $r_s=0.5$, and our choice of $\gamma$ implies that $\rho\propto r^{-2}$. All other quantities are derived from the combination of these 
parameters with a given value for $f_p$. In \cref{fig:profiles}, we plot the velocity and pressure profiles for our chosen values of $f_p$ that we will explore in this paper. From 
the figure it is clear that for higher values of $f_p$, the pressure support increases while the support from centripetal acceleration decreases.

In the absence of friction, each disk profile should remain unaltered over time from the initial conditions plotted in \cref{fig:profiles}. 
In the next section, we demonstrate the efficacy of several artificial viscosity methods at achieving this result.

\section{Results}
\label{sec:results}
\begin{figure}[ht]
\centering
\includegraphics[width=0.45\textwidth]{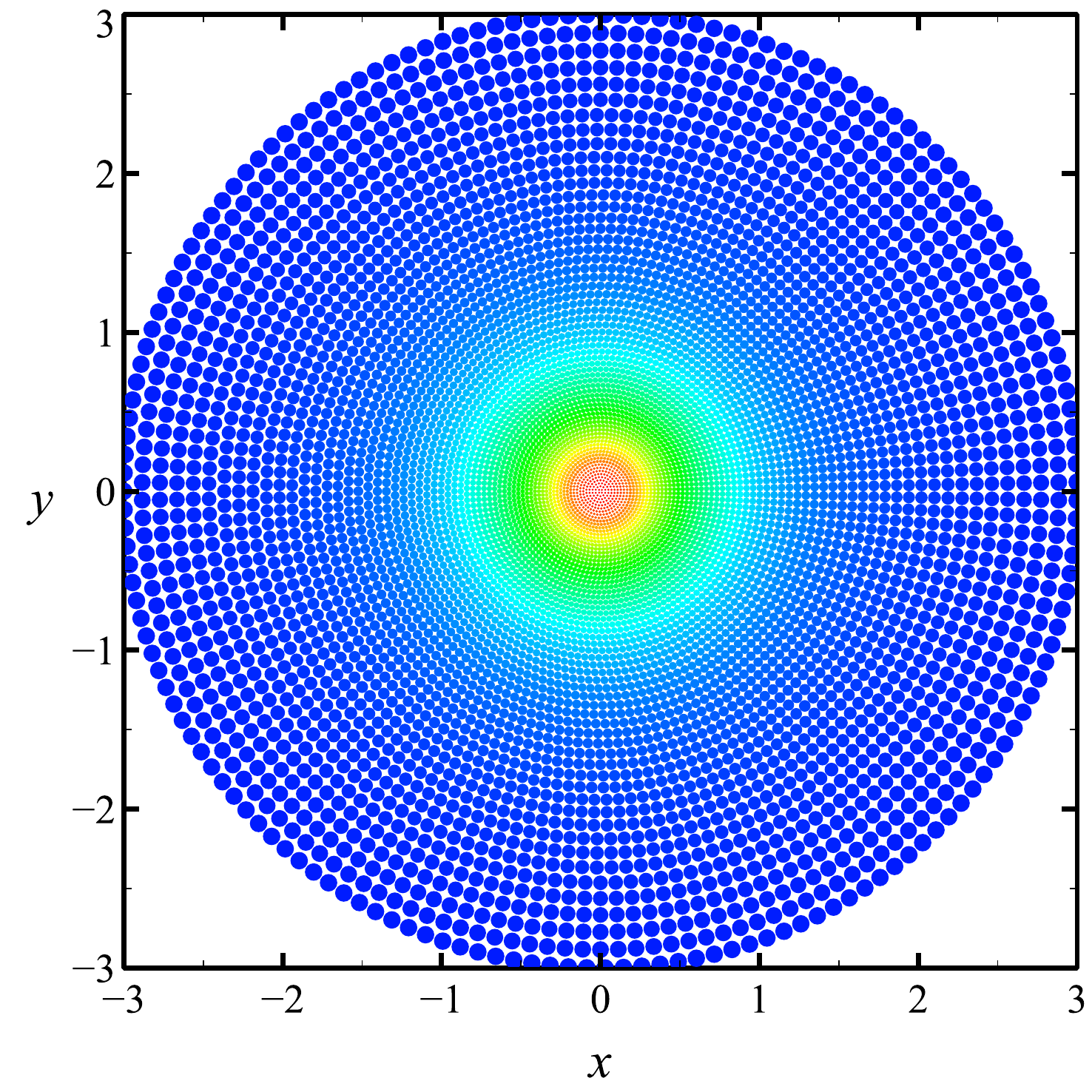}
\caption{An initial arrangement of particles for an example simulation using the constant-$\delta\theta_r$ setup. Particles are colored by density and scaled by their radial coordinate.}
\label{fig:setup}
\end{figure}

The results presented here use 2D geometry to constrain particle motions within the $x-y$ plane. 
However, as our central potential follows the familiar $\Phi \propto r^{-1}$ law, and since we neglect self-gravity, these disks are analogous to razor-thin disks in 3D geometry.
For each value of $f_p$, we arrange the particles in the plane with 50 radial annuli of particles, all of equal mass. 
This results in $\approx7800$ total particles in each simulation. 
To match the analytical density profiles, the angular and radial separations between particles are adjusted in each annulus such that the angular separation between particles at a given radial coordinate ($\delta\theta_r$) is a constant \citep{owen1998,cartwright2009}.
An example initial setup is shown in \cref{fig:setup}.

For all of our simulations, we employ the compatible energy differencing evolution described in \cite{owen2014}. 
This form of energy evolution has been shown to conserve energy to machine precision, and while traditional SPH absent this modification conserves momentum explicitly, energy conservation is less precise.
We test three variants of SPH with different prescriptions for artificial viscosity -- M\&G, Balsara, and C\&D viscosity -- as well as CRKSPH with the Christensen-like viscosity. 
For the three variants of traditional SPH, we use a fifth-order B-Spline kernel \citep{Schoenberg1972}, whereas for CRKSPH we employ the seventh-order B-Spline.
Each of these choices is made based on experimenting with what works optimially for each method.
It is well-known that SPH benefits from the fifth-order spline in order to resist local particle clumping.
The seventh-order spline choice used for CRKSPH is consistent with the choice made in \cite{frontiere2016}: CRKSPH is relatively insensitive the details of this choice (fifth vs.~seventh-order), but it was found that results with the seventh-order kernel were slightly better.
Regardless of kernel choice, in all cases we adjust the smoothing scales so each point maintains roughly a radial sampling of 4 neighbors, implying in 2D a total of $\approx 48$ neighbors per point.
Finally, we explore results using these modeling techniques for $f_p \in \{0.05,0.50,0.95\}$.

\subsection{A Low Pressure Disk}

\begin{figure}[ht]
\centering
\includegraphics[width=0.40\textwidth]{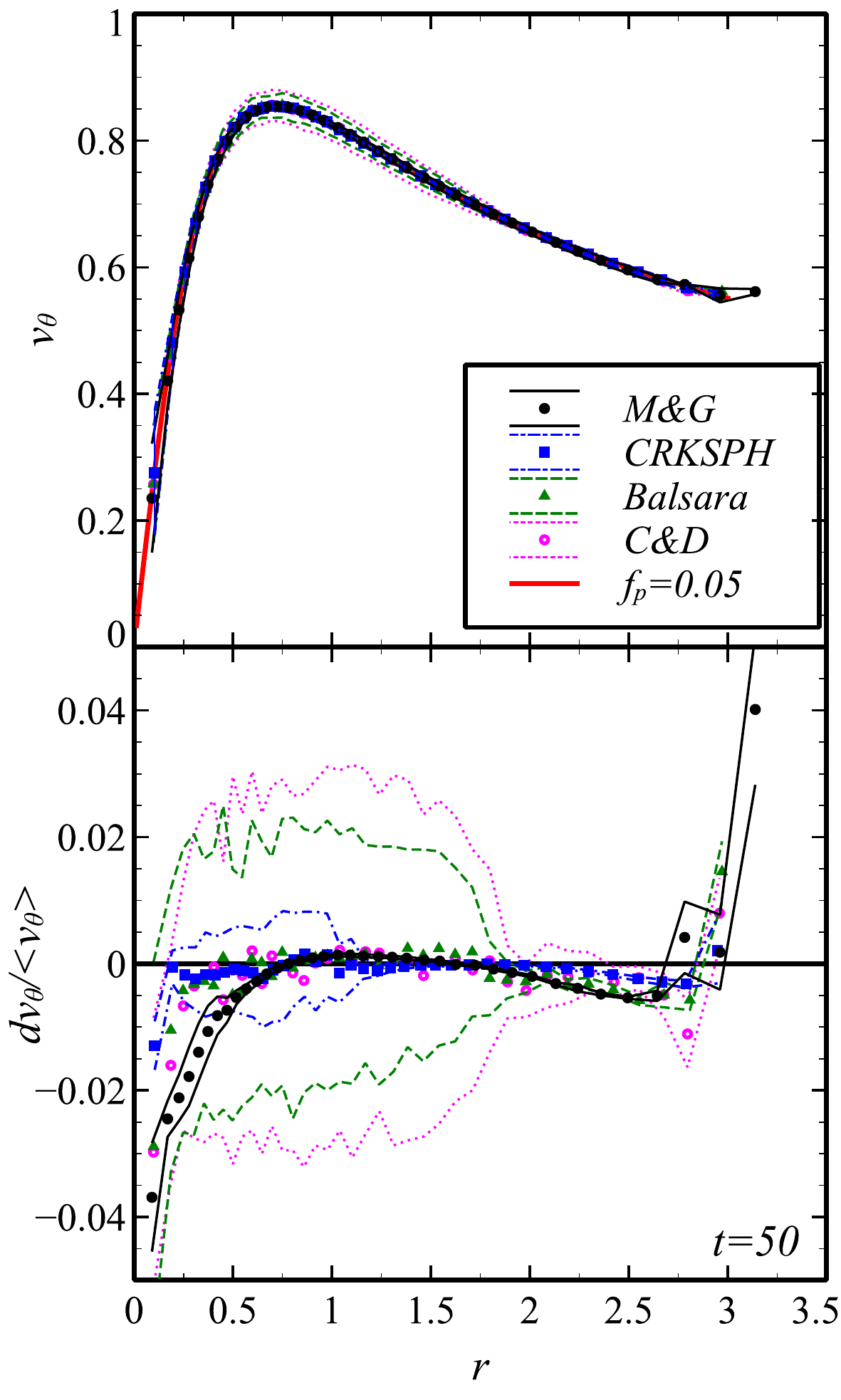}
\caption{Top: Radially binned (200 particles per bin) velocity profiles for each of our $f_p=0.05$ simulations at $t=50$. The standard deviations in each bin are indicated by the curves bracketing each data sample. Bottom: Fractional errors from the analytical expectation for the same binned data.}
\label{fig:0p05v}
\end{figure}

\begin{figure}[ht]
\centering
\includegraphics[width=0.45\textwidth]{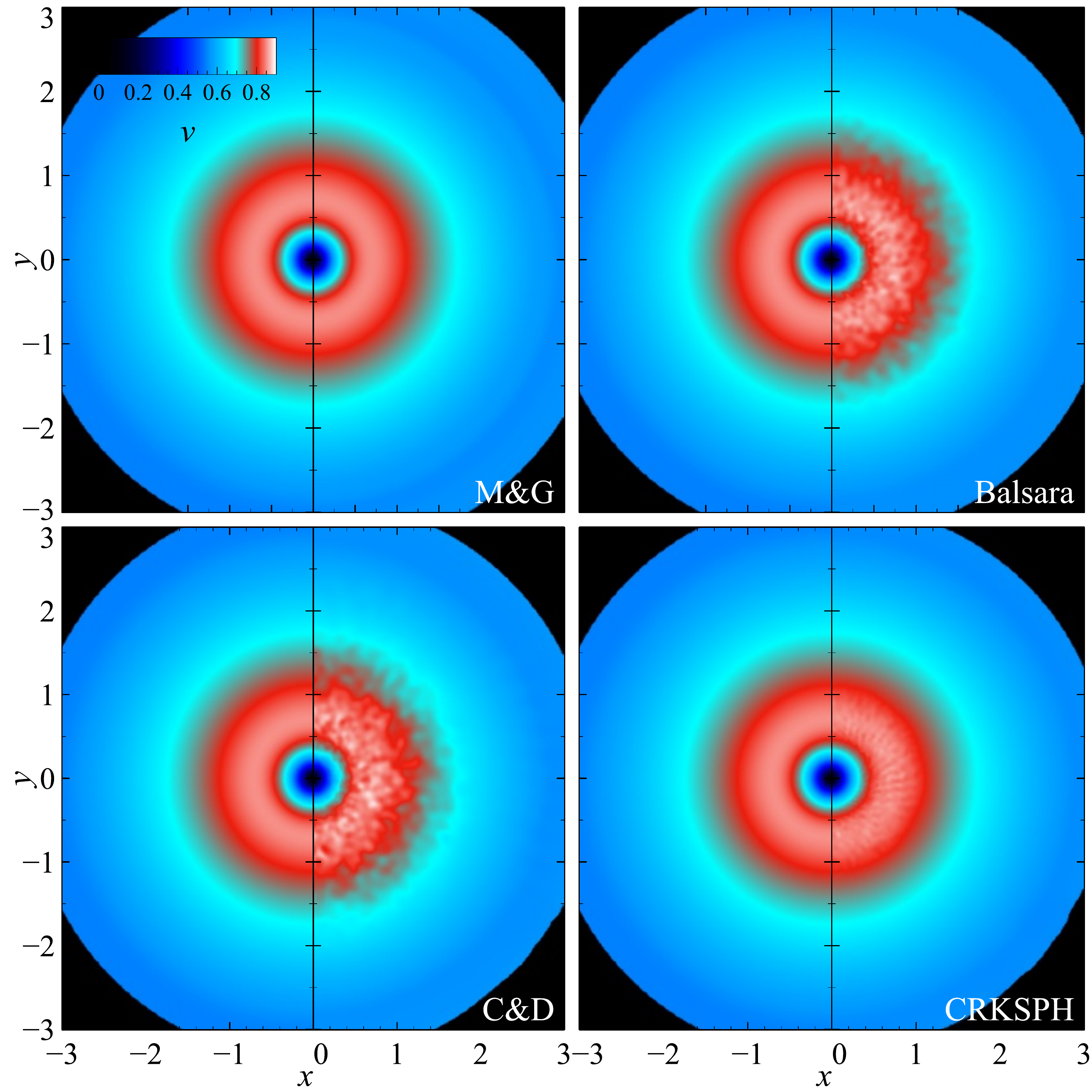}
\caption{2D velocity maps for each $f_p=0.05$ simulation at $t=50$, with the left half of each panel showing the initial condition for reference.}
\label{fig:0p05map}
\end{figure}

First, we examine the low-pressure case where $f_p=0.05$.
In \cref{fig:0p05v}, we plot the velocity profiles of each of our simulations at a time $t=50$ for $f_p=0.05$ with radial bins containing 200 particles each, as well as the deviation from the initial condition. 
\Cref{fig:0p05map} shows images of the 2D velocity fields for the same $t$ and $f_p$.  At this time, the particles near the peak of the velocity curve (see \cref{fig:profiles}) have orbited roughly 10 times.
This scenario is most similar to pressure-free tests often performed in the literature, but still features just enough pressure such that the proper treatment would not involve the trivialization of hydrodynamics.
Physically, this resembles the balance typical of proto-planetary disks \citep{lissauer1987,lubow1999,boss2001}, where much of the material is cold, but gas dynamics still play a minor role.

In this case, the standard M\&G viscosity appears to perform quite well in the 2D maps, but the binned radial profiles reveal larger systematic errors in the momentum near the center and the edge of the simulations.
The Balsara and C\&D simulations have less systematic error, but a much larger scatter in their velocities. 
This is most evident in the 2D maps of \cref{fig:0p05map}.
CRKSPH, on the other hand, has less overall scatter than either C\&D or Balsara, while also featuring less systematic error than any of the other methods.

\begin{figure}[ht]
\centering
\includegraphics[width=0.40\textwidth]{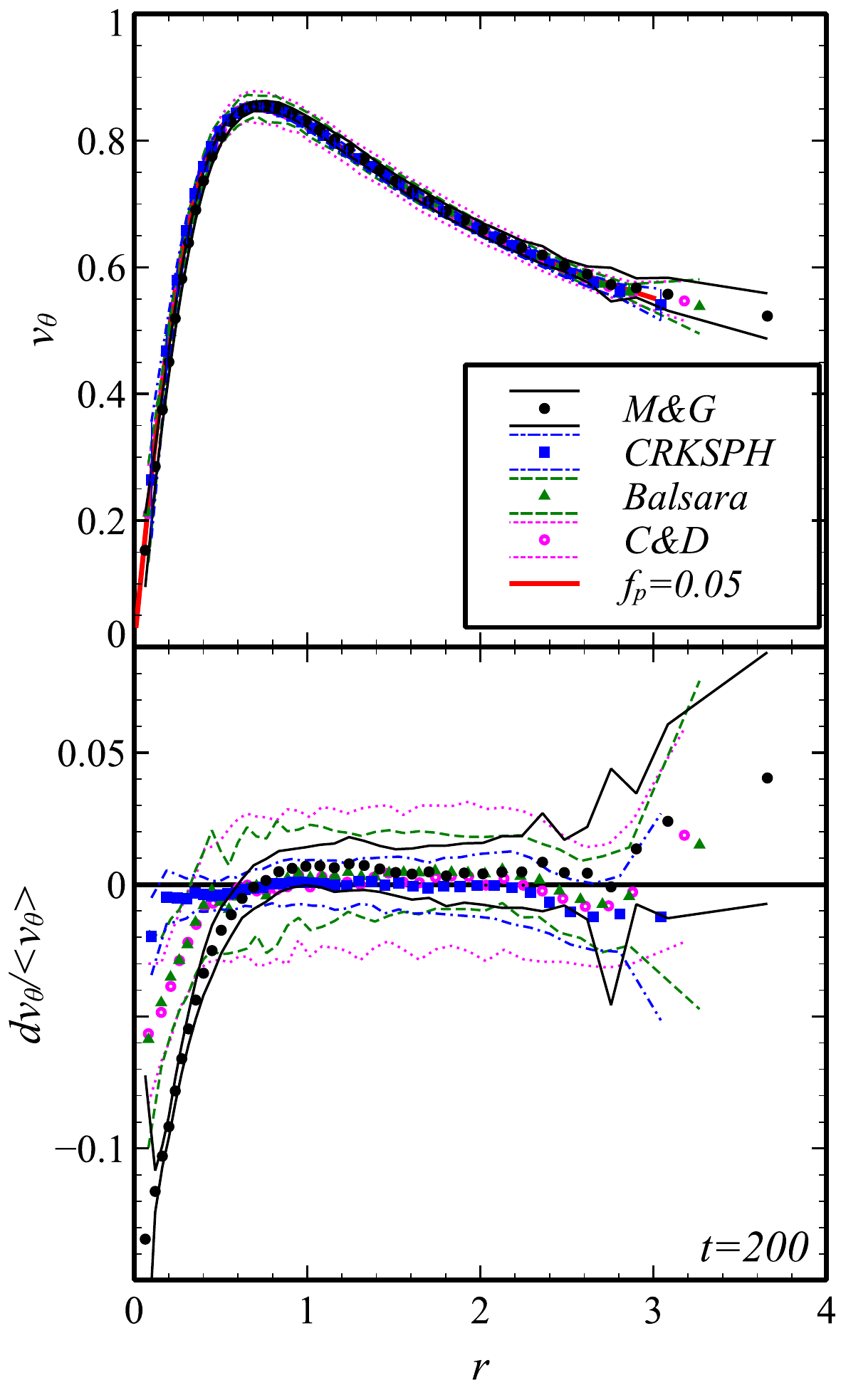}
\caption{Same as \cref{fig:0p05v} but for $t=200$.}
\label{fig:0p05v200t}
\end{figure}

\begin{figure}[ht]
\centering
\includegraphics[width=0.45\textwidth]{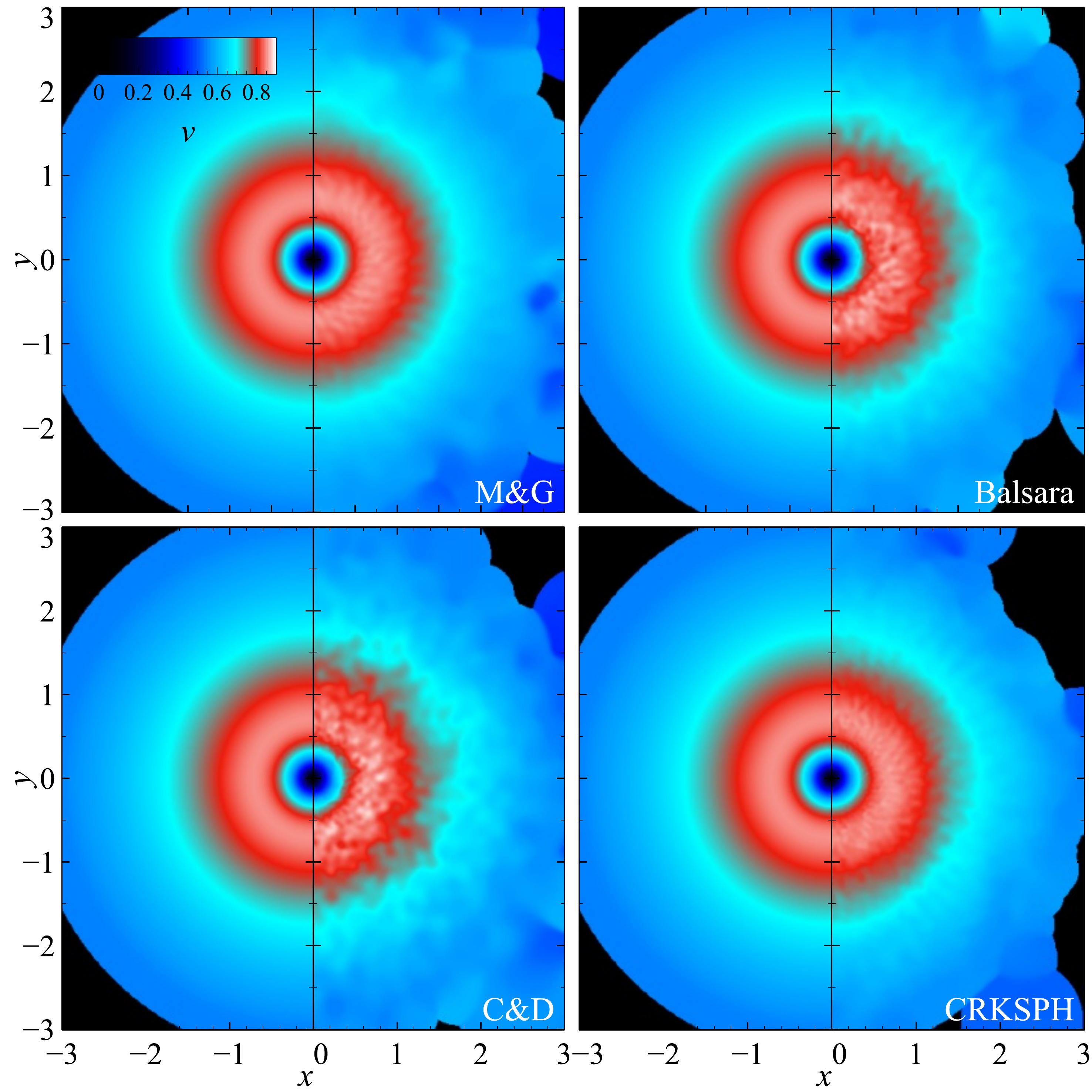}
\caption{Same as \cref{fig:0p05map} but for $t=200$.}
\label{fig:0p05map200t}
\end{figure}

If we continue the $f_p=0.05$ simulations out to $t=200$ (roughly 40 orbits at peak velocity), the systematic errors grow over time as shown in \cref{fig:0p05v200t}.
Whereas the M\&G simulation had accumulated $\approx3\%$ error in the rotational velocity near the center at $t=50$, by $t=200$, this error has grown to $\gtrsim10\%$.
Balsara and C\&D results similarly show a systematic drift toward larger errors near the center.
CRKSPH, however, appears relatively unchanged from the $t=50$ result, indicating that resolution near the center is the dominant error here for this method.
All of these methods have begun to cast particles out of the disk by this time, though CRKSPH has the least mass loss of the four methods tested, as shown in \cref{fig:0p05map200t}.

\begin{figure}[ht]
\centering
\includegraphics[width=0.45\textwidth]{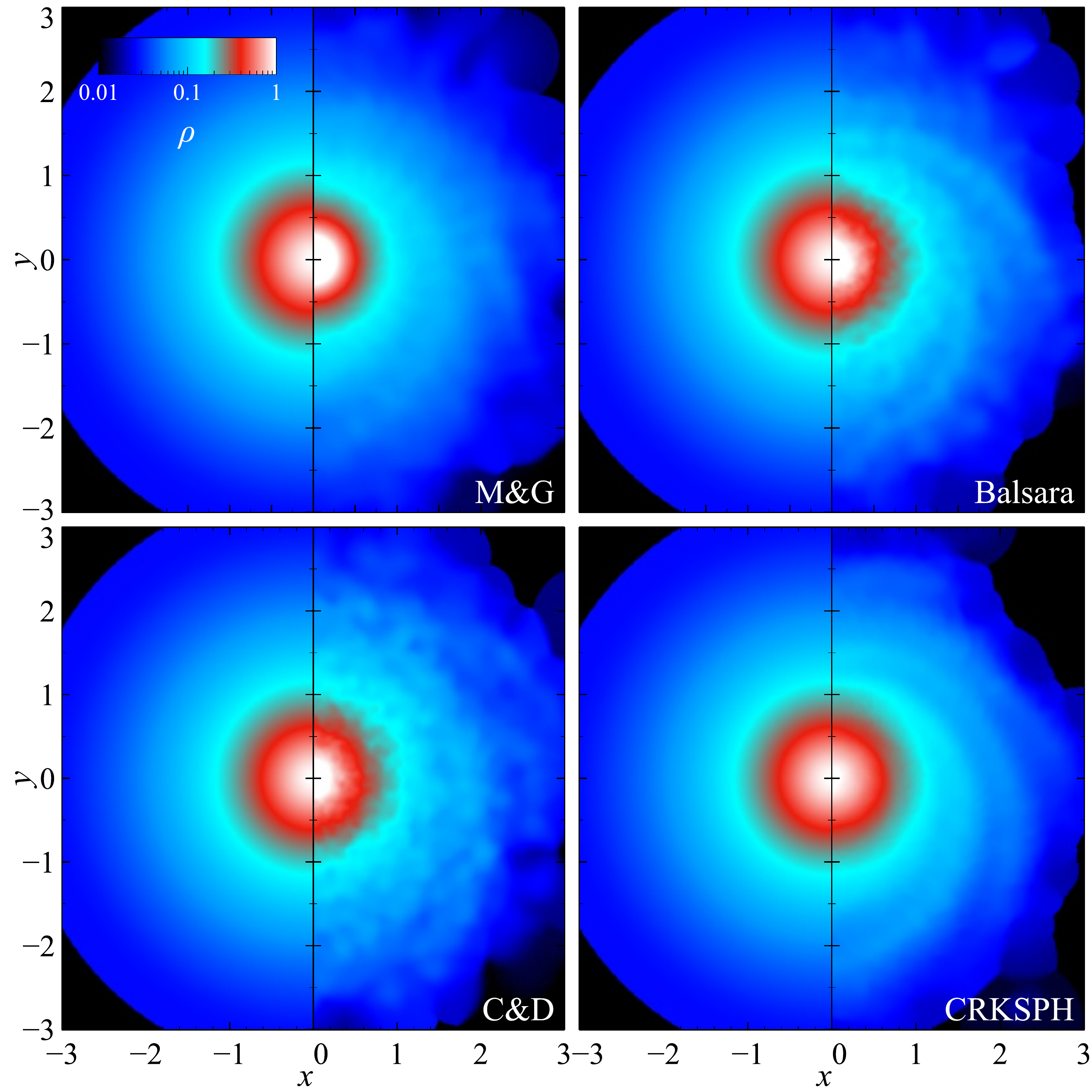}
\caption{2D density maps for each $f_p=0.05$ simulation at $t=200$, with the left half of each panel showing the initial condition for reference.}
\label{fig:0p05denmap}
\end{figure}

If we examine the density maps of these disks at this late time, shown in \cref{fig:0p05denmap}, -- specifically the radial location of the red contour -- it is clear that the disk is collapsing in all three of the SPH realizations to varying degrees as the momentum support has been transported away from the center of the disk.
Only the CRKSPH simulation has a more-or-less unperturbed density profile from the initial condition, however, as this is a low-pressure simulation, the evolution in each realization is very slight.

\subsection{A High Pressure Disk}

\begin{figure}[ht]
\centering
\includegraphics[width=0.40\textwidth]{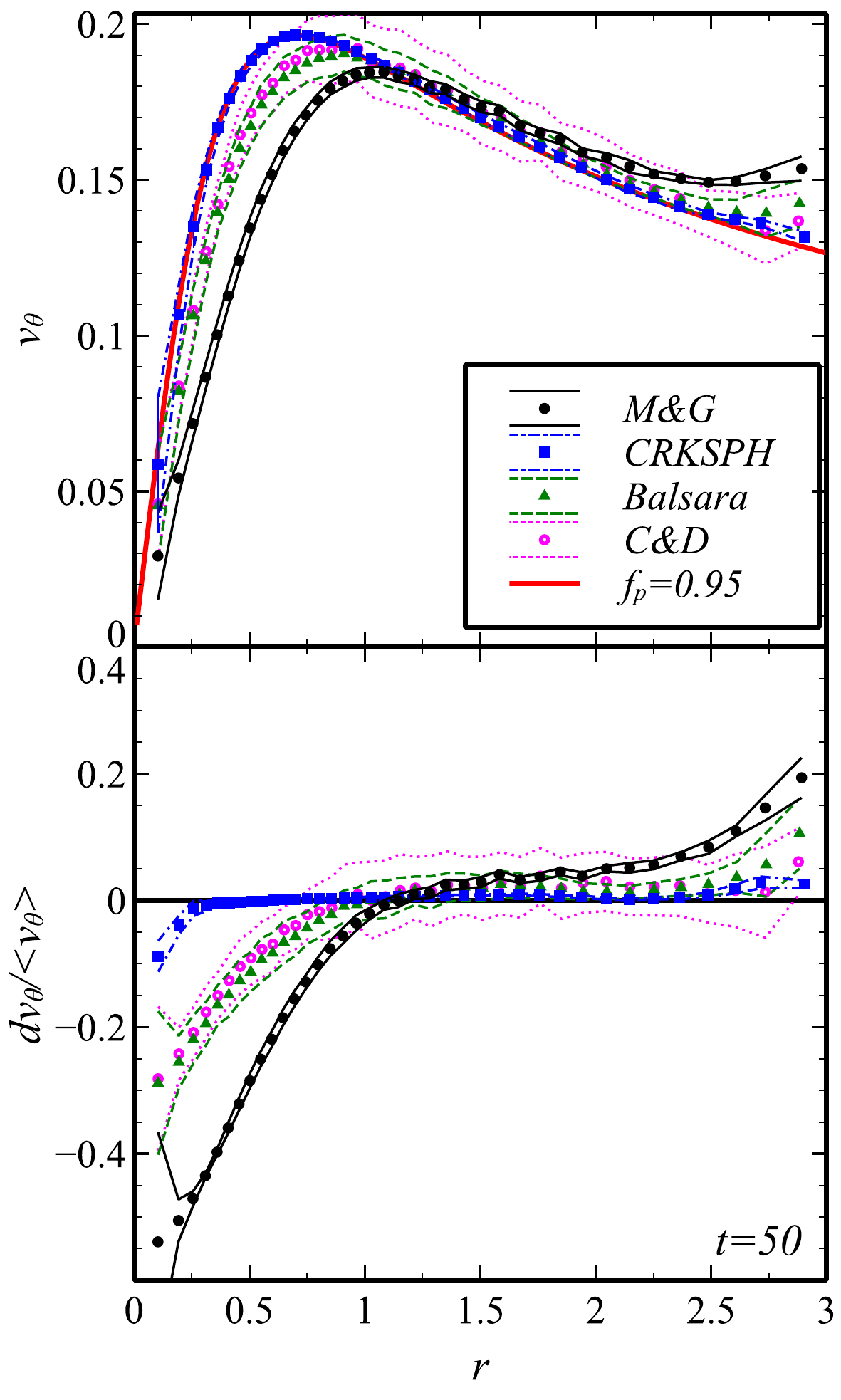}
\caption{Top: Radially binned (200 particles per bin) velocity profiles for each of our $f_p=0.95$ simulations at $t=50$. The standard deviations in each bin are indicated by the curves bracketing each data sample. Bottom: Fractional errors from the analytical expectation for the same binned data.}
\label{fig:0p95v}
\end{figure}

\begin{figure}[ht]
\centering
\includegraphics[width=0.45\textwidth]{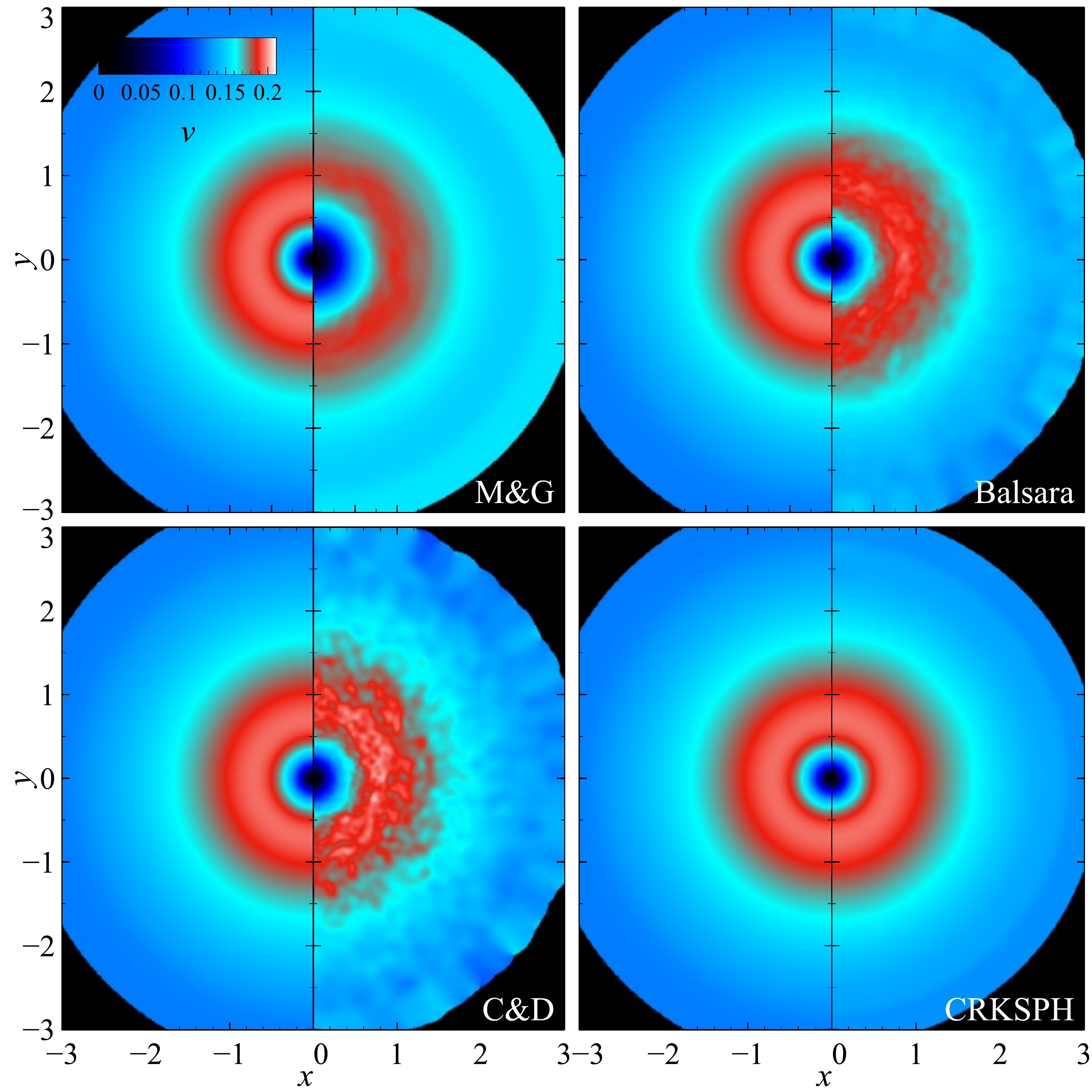}
\caption{2D velocity maps for each $f_p=0.95$ simulation at $t=50$, with the left half of each panel showing the initial condition for reference.}
\label{fig:0p95map}
\end{figure}

Next we turn our attention to the opposite extreme: a disk dominated by pressure support with $f_p=0.95$.
\Crefrange{fig:0p95v}{fig:0p95map} show the resulting radial profiles and velocity images at $t=50$: a similar dynamical time to our $t=50$ example in the low-pressure support case, but in this case the peak velocity points have only orbited roughly twice.
This is a high value for the pressure support and provides a very different test of our candidate methods.
%% In this situation the sound speed is large, so while the quadratic term in the viscosity (\cref{eq:MGPi}) still scales $\propto \Delta v^2$, the linear term is no longer negligible as it scales like the sound speed times the pair-wise velocity difference ($\propto c_s \Delta v$).  As the sound speed grows this linear viscosity term can dominate the overall viscosity contribution.
%%This is a high value for the pressure support and provides a very different test of artificial viscosity.
%%In this situation the sound speed is large, so while the quadratic term in the viscosity (\cref{eq:MGPi}) still scales $\propto \Delta v^2$, the linear term is no longer negligible as it scales like the sound speed times the pair-wise velocity difference ($\propto c_s \Delta v$).  As the sound speed grows this linear viscosity term can dominate the overall viscosity contribution.

As the figures demonstrate, SPH using the unmodified M\&G viscosity fares the poorest for this test, with the largest absolute error at low and high radial bins.
The Balsara correction and C\&D viscosity are competitive with one another, though the C\&D viscosity results in a larger scatter. 
This is likely due to the rapid activation and deactivation of the viscosity as the limiter function described in \crefrange{eq:cullen}{eq:vsig} is not able to fully disentangle shearing motion from convergent flows. 
The Balsara switch suffers from the same limitation, but as it is merely a multiplier on the M\&G viscosity that smoothly changes as the fluid evolves, the total viscosity is less susceptible to rapid fluctuations compared with the C\&D viscosity.
The CRKSPH result is very nearly unaltered from the initial condition, owing to a combination of a more accurate discretization and effective limiting of artificial viscosity.

The velocity field maps of \cref{fig:0p95map} reveal a similar pattern. 
The M\&G viscosity simulation shows the most deviation from the initial condition, while the Balsara and C\&D viscosity simulations feature less deviation but more scatter. The CRKSPH field map appears almost unchanged from the initial condition. 

The errors seen in the three SPH realizations are not trivial. They represent wholesale momentum transport from the inner regions of the disk to the outer regions. 
As our setup does not include any instabilities that should drive momentum to higher radii (such as magneto-rotational instabilities), this result could easily be confused with an actual physical process, when in actuality, this is merely a numerical error.

\begin{figure}[ht]
\centering
\includegraphics[width=0.40\textwidth]{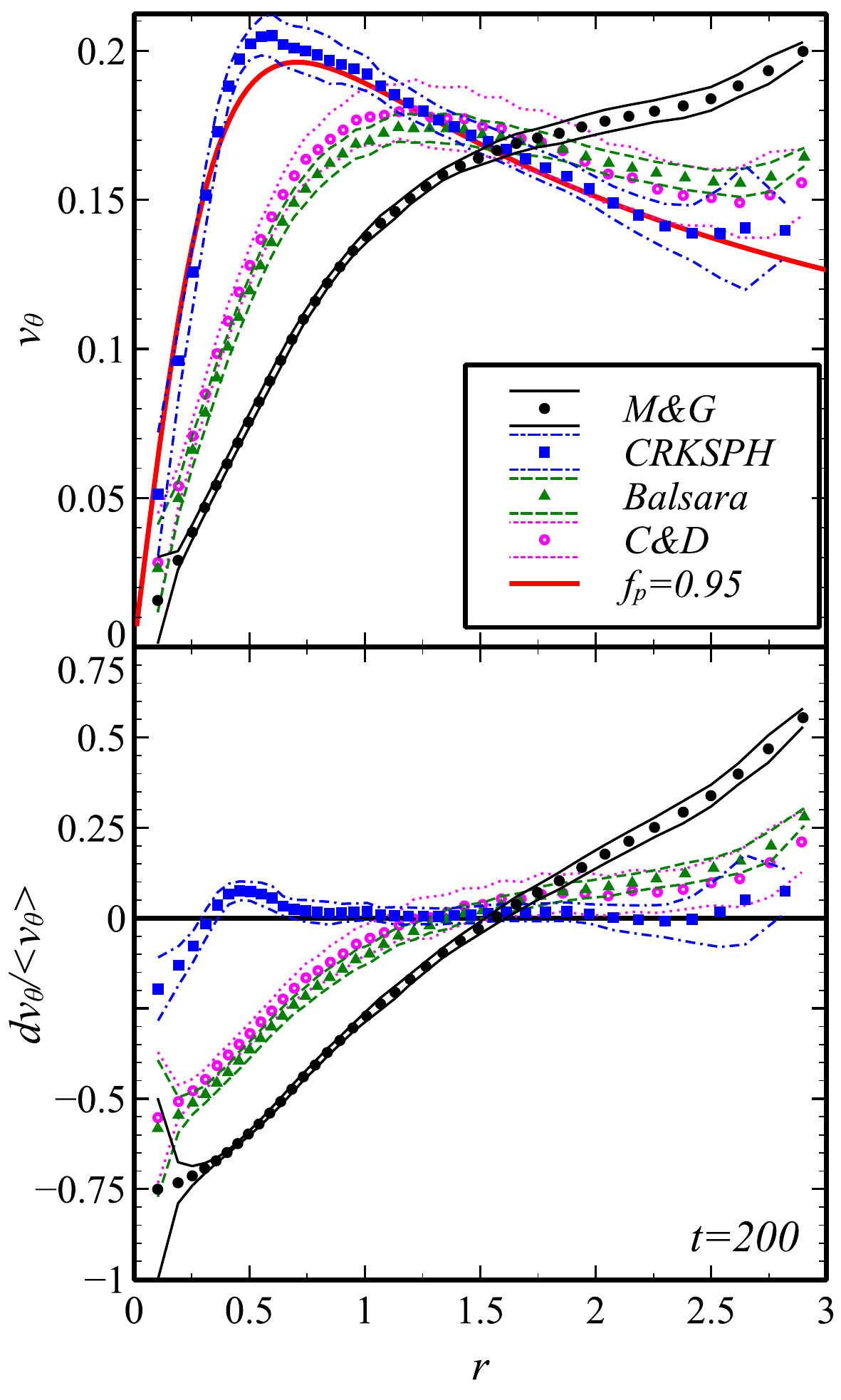}
\caption{Same as \cref{fig:0p95v} but for $t=200$.}
\label{fig:0p95v200t}
\end{figure}

If we continue the $f_p=0.95$ simulations out to $t=200$ (as we did with the $f_p=0.05$ simulations), the systematic errors grow over time, as seen in \cref{fig:0p95v200t}.
This time represents just over 7 orbits for the nominal peak velocity radius.
The M\&G simulation no longer resembles the initial condition velocity curve, and the errors near the center and edge have grown to $\approx75\%$. 
This is evidence of large-scale momentum transport to the outer edges of the disk.
The Balsara and C\&D realizations have also drifted quite considerably from the initial condition, with velocity errors of $\approx50\%$ near the center.
CRKSPH on the other hand, has maintained a much closer fit to the initial setup, even at $t=200$, with most of the error appearing inside the softening length of the gravitational potential.

\begin{figure}[ht]
\centering
\includegraphics[width=0.40\textwidth]{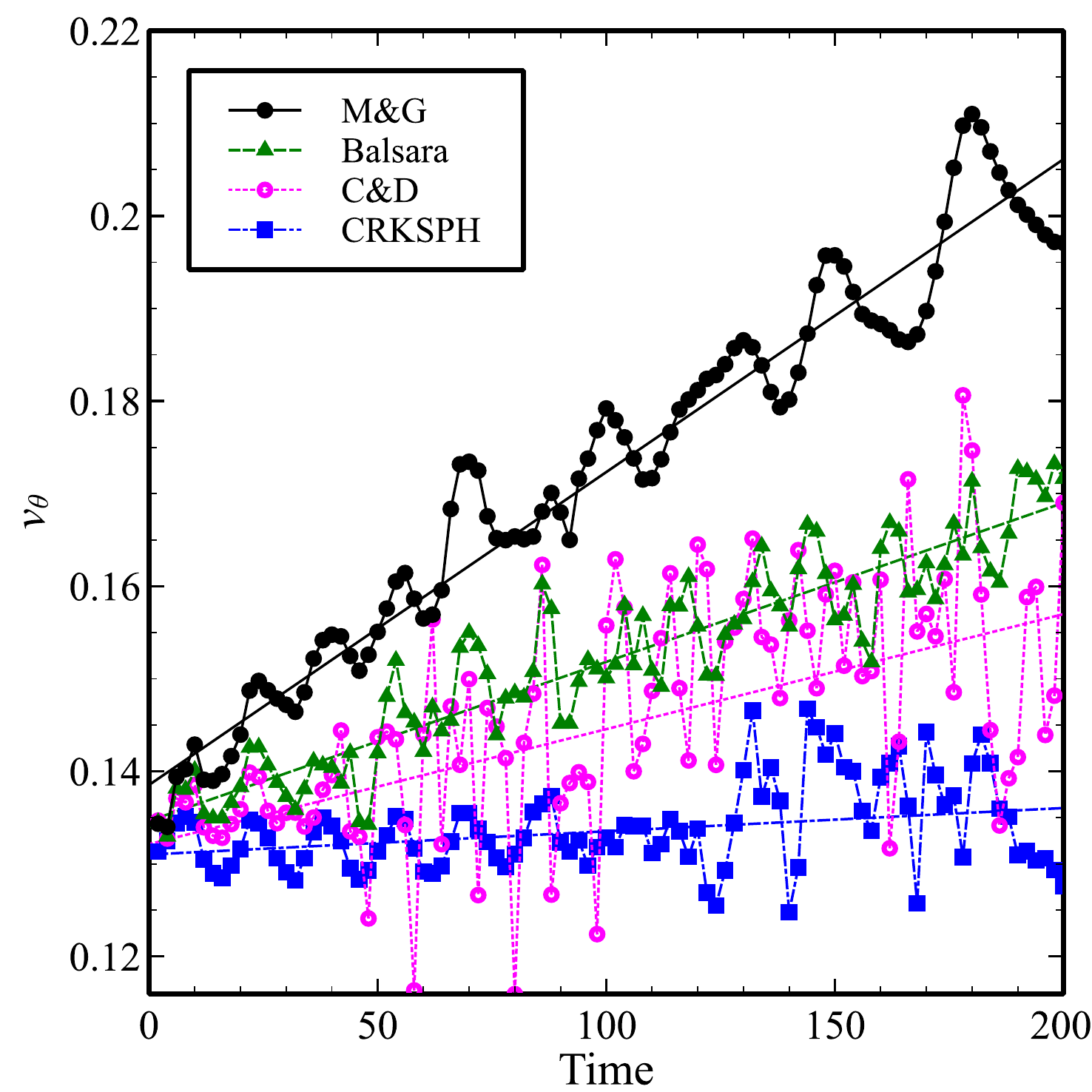}
\caption{The evolution over time of the azimuthal velocity of a single particle initially at $r\approx3$ in each of the $f_p=0.95$ realizations. The high-frequency oscillations of period $t \approx 20$ is due to the simulation relaxing toward equilibrium due to imperfections in the initial conditions, described in the text.}
\label{fig:0p95evo}
\end{figure}

In \cref{fig:0p95evo}, we plot the velocity history of a random particle in the outermost radial bin of each simulation at $f_p=0.95$. 
From the slopes of the velocity curves, we can estimate an approximate numerical acceleration that particles near the edge experience.
This corresponds directly to a pseudo-force that arises from improper activation of the artificial viscosity and any inaccuracies in the hydrodynamical forces.
The oscillations seen in the figure are due to the imperfect numerical representation of the continuum equilibrium condition: a combination of the fact the disk is terminated at a finite radius and the summation definition of the mass density deviating from the ideal desired expection.
As a result, there is some error in the calculation of this acceleration term, and in \cref{table:0p95accel}, we give the accelerations ($a$) and errors ($\delta a$) for each of these simulations.
In other words, the acceleration $a$ is the linear regression slope fit of the histories in \cref{fig:0p95evo} and $\delta a$ is the standard error for these regression fits.
The pseudo-force in CRKSPH is quite small and of the same order as the error for the estimate, which is desirable since the physically motivated value for $a$ is zero.

\begin{table}[ht]
\caption{The erroneous accelerations ($a$) and errors ($\delta a$) for a single particle chosen from each realization of the $f_p=0.95$ disk, corresponding to the histories depicted graphically in \cref{fig:0p95evo}. }
\centering
\begin{tabular}{l | c c}
\hline\hline
 & $a$ [$\times10^-5$] & $\delta a$ [$\times10^-5$]\\
\hline
M\&G	& 	33.8 & 	1.46 \\
Balsara	& 	17.2 & 	1.31 \\
C\&D	& 	12.4 & 	1.39 \\
CRKSPH	& 	2.51 & 	1.14 \\
\hline
\end{tabular}
\label{table:0p95accel}
\end{table}

At high pressures such as these, removing support against gravity via momentum transport merely serves to rebalance the disk closer to full pressure support.
As a result, the density curves do not appear drastically altered from the initial condition, though in the three SPH realizations, there is a slight over-density ($\rho>1$) at the very center of the disk.

\subsection{Sound Speed and Artificial Viscosity}

At first blush it might seem a surprising result that the disk dominated by pressure support ($f_p=0.95$) shows much larger errors in the rotational curves than the case dominated by rotational support ($f_p=0.05$).
In each of these models the dominant error term arises from improper activation of the artificial viscosity, which is proportional to the velocity difference between interacting points, so why does the error grow so much more dramatically in the case with lower rotational velocities?

The answer lies in the formulation of the various terms that contribute to the viscosity.
If we examine the definition of the classic M\&G viscosity in \cref{eq:MGPi} we see there are two terms: the quadratic term which scales like the relative velocity jump between each interacting pair of points, $\Piquad \propto (\Delta v)^2$, and a linear term that scales like the sound speed multiplied by the velocity jump, $\Pilin \propto c_s \Delta v$.
For the low-pressure case ($f_p \ll 1$) the linear term is suppressed since the sound speed becomes negligible.
In this situation, only the quadratic term contributes to the viscosity, and in these rotational shearing flows, the relative velocity jump between points in similar orbits is relatively small, evident in the rotational curves of \cref{fig:profiles}.
This limit is very close to the classical pressureless Keplerian disk, and therefore it is not surprising that our $f_p=0.05$ results look quite similar to those found in previous investigations \citep{cullen2010,cartwright2010,hosono2016}.

However, for the high-pressure case as $f_p \to 1$, the linear term of the viscosity begins to dominate.
This is because while the velocity jump between points on neighboring orbits may be relatively small, the sound speed becomes large so $c_s \Delta v > (\Delta v)^2$.
This larger contribution to the viscosity from the linear term tips the scales and allows the artificial viscosity to become more destructive in the pressure dominated case even though the orbital velocities are slower than those found in the rotationally dominated disk.

The effects of the viscosity on both classes of problems are the same, it is just the magnitude of the error that is affected by the degree of pressure support.
In the M\&G simulation this error is essentially constant, yielding a monotonic divergence from the initial condition over time.
For Balsara and C\&D, the viscosity scaling correction is allowed to flucutate relatively rapidly, resulting in a larger scatter, but less overall error than the M\&G simulation.
The limited viscosity in CRKSPH fares the best in these tests, limiting the viscosity effectively without introducing too much scatter in the curves.

\subsection{An Intermediate Disk}

Finally, we explore an intermediate case between the previous extremes with $f_p=0.50$.
For accretion disks around compact objects, pressure and rotation play an equal role \citep{rosswog2009wd,rosswog2010,vanKerkwijk2010,raskin2010,raskin2012,raskin2014}.
There, the high temperatures ($T\sim10^8$ K) and densities ($\rho\sim10^5$ g cm${}^{-3}$) result in $\approx50-75\%$ pressure support, and this is most easily demonstrated by the rotational plots of \cite{raskin2012} indicating sub-keplerian orbital velocities.
These conditions roughly equate to our idealized problem here with $f_p=0.50$ -- studying our models with this level of pressure support should serve as a useful probe of how applicable each simulation technique might be to this real-world interesting case.

\begin{figure}[ht]
\centering
\includegraphics[width=0.40\textwidth]{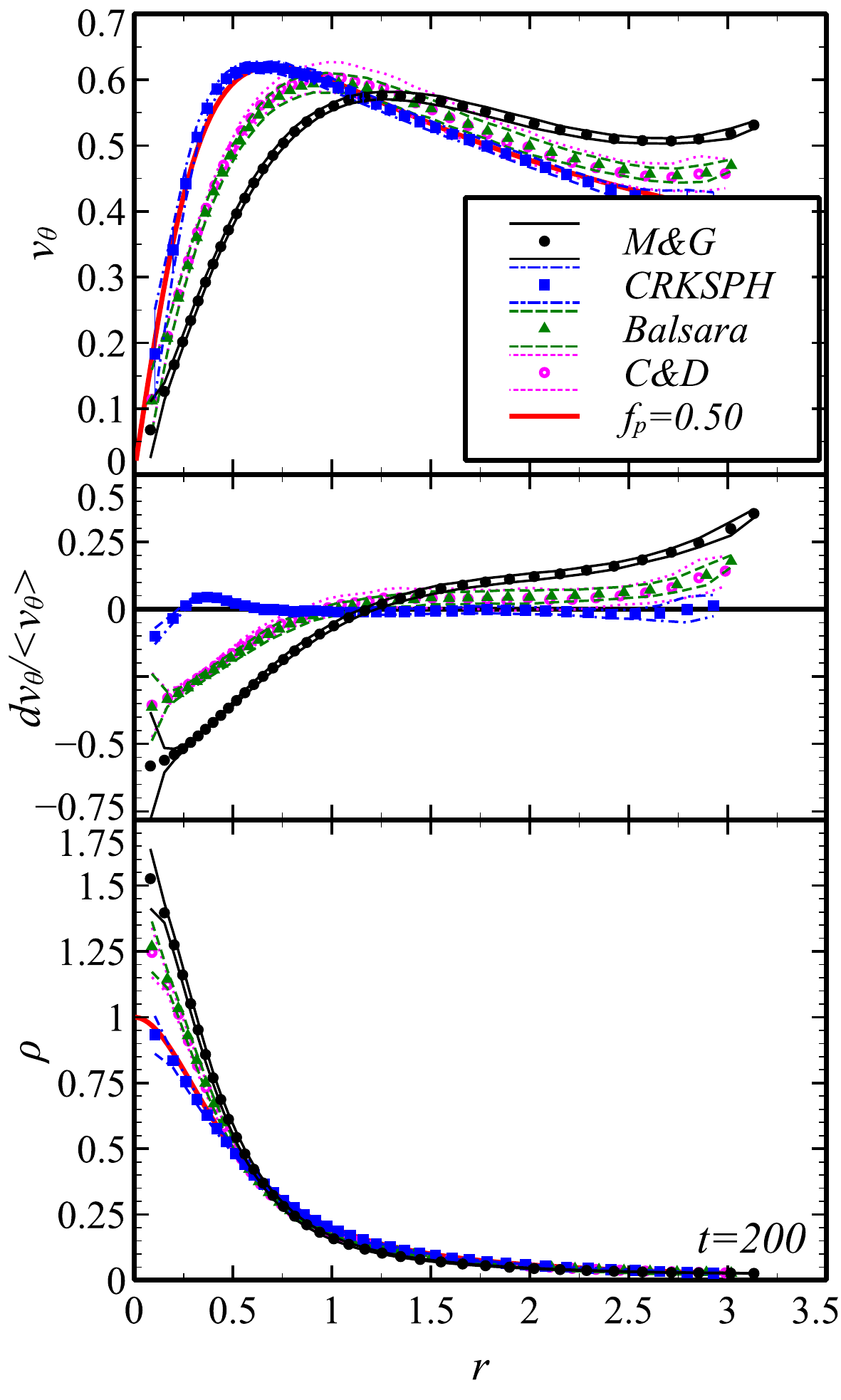}
\caption{Top: Radially binned (200 particles per bin) velocity profiles for each of our $f_p=0.50$ simulations at $t=200$. The standard deviations in each bin are indicated by the curves bracketing each data sample. Middle: Fractional errors from the analytical expectation for the same binned data. Bottom: Density profiles for the same binned data.}
\label{fig:0p50v200t}
\end{figure}

In \cref{fig:0p50v200t}, we plot radial profiles of the velocity and density for models using $f_p=0.50$ at $t=200$.
These results are quite similar to those of the $f_p=0.95$ simulations in that the M\&G simulation has largest deviation and least scatter.
Balsara and C\&D again have a very comparable result with moderate deviation and a large scatter.
The CRKSPH simulation performs the best of the four approaches, with only minimal deviation near the center and edge. 

In the case of a real compact object accretion disk application, the M\&G, Balsara, and C\&D approaches would all result in unphysical momentum transport away from the compact object after only a handful of orbits. 
Crucially, the central over-densities seen in the bottom panel of \cref{fig:0p50v200t}, with M\&G as high as 75\% over-dense, are a drastic departure from the initial condition, and could result in disk material reaching nuclear-burning conditions in a real application, where in reality, this material should be quiescent.

\subsection{Disk Lifetimes}

For each choice of the fractional pressure support ($f_p$), the erroneous angular momentum transport demonstrated in Figure \ref{fig:0p95evo} and enumerated in table \ref{table:0p95accel} trends the disk material toward an equilibrium solution of increasing pressure support at the cost of rotational support. 
For many previously explored astrophysical disk tests \citep[\eg][]{hosono2016} where a strict inner boundary was enforced with a voided region (typically to avoid singular forces from an unsoftened potential, but also to more accurately approximate protoplanetary disks), this evolution would cause catastrophic collapse of the simulated disks on finite time scales.
In our case, our disk simulations continue to evolve essentially forever due to the combination of a softened potential and the elimination of the assumption of zero pressure. 
Over very long timescales, we can expect all of our simulations to evolve toward purely pressure supported, static disks as the various artificial viscosity treatments convert kinetic energy into thermal energy.

\subsection{Contribution to Errors}

As CRKSPH has outperformed SPH with various prescriptions for artificial viscosity in each of the tests in this examination, a question naturally arises as to the relative importance of errors from the SPH method itself as compared to errors arising from the artificial viscosity treatment. 
In both cases, CRKSPH and SPH, there is an unavoidable error in the measurement of the pressure gradient that stems mainly from the way that both methods construct neighbor sets that spatially overlap many annuli with differing rotational velocities and pressures. 
This is most crucially in error near the center of the disk, where a particle's neighbor set will include particles on the \textit{opposite side of the disk}, whose velocities and pressure gradients are pointing in opposite directions.

As the pressure gradient in any SPH prescription is a smoothed quantity over the local neighbor set, the derived, pair-wise force between particles from pressure will feature this unavoidable error that grows in magnitude at increasingly smaller radii.
CRKSPH is not immune to this error \citep{frontiere2016}, and indeed, when we measure the instantaneous pressure gradient at $t=0$ and compare to the analytical expectation, both CRKSPH and SPH have essentially the same error. 
However, this kind of measurement is really only feasible for $t=0$ since in all cases, the disk immediately evolves away from the analytical expectation, and any error measurement in the pressure gradient becomes incredibly difficult to tease out of the numerics. 

What we can say with some confidence is that this error is both small and non-uniform.
While the artificial viscosity errors (recalling that \textit{any} viscosity is an error in this problem) are unidirectional in magnitude and continuously compound to evolve the disks away from their initial conditions, the error in the pressure gradient is a universal and numerical mismatch of the underlying numerics to the expectation from continuum fluid mechanics.
As a result, force errors from this mismatch can point in truly any direction, and so the net effect is essentially negligible. 

Far more important is the artificial viscosity treatment, and this is borne out from the results of the different values for $f_p$ we tested here. 
The error in the measurement of $\nabla P$ is not proportional to $P$, and so will be essentially a constant for all values of $f_p$. 
In fact, the error in $\nabla P$ is really only proportional to our spatial resolution.
When looking at a single realization of SPH, \eg the M\&G curves in Figures \ref{fig:0p05v200t} \& \ref{fig:0p95v200t}, there is a dramatic difference in the evolution of the disk with different values for $f_p$, much more so than there is between CRKSPH and SPH of any flavor when artificial viscosity is more limited, as in Figure \ref{fig:0p05v200t}. 

\section{Discussion}
\label{sec:conclusion}
In this paper, we have developed a simple test problem to appraise the treatment of hydrodynamics and gravitation for a realistic, rotating disk simulation based on the work of \cite{owen1998}.
We believe this prescription is a more appropriate test of the performance of any hydrodynamical method for applications modeling astrophysical disks as compared with the ordinary, purely gravitational Keplerian disk.
We have tested the performance of three popular SPH formulations: one using the ordinary M\&G viscosity, one employing the Balsara viscosity multiplier, and one using the more sophisticated C\&D viscosity formulation.
We also examined the performance of a new meshfree method based on SPH: CRKSPH \citep{frontiere2016}.
For each choice of the relative ratio of pressure to rotational support ($f_p$), CRKSPH outperformed all other methods with the least systematic error and spurious momentum transport.
The addition of pressure support into the treatment of a (partially) rotationally supported disk reveals many weaknesses in some popular flavors of artificial viscosity developed for SPH.
When the linear term of the artificial viscosity is not ameliorated via a negligible sound speed, the ability of the artificial viscosity scheme to distinguish shear from compression becomes the central governing mechanism by which the disk evolves. 

While these specific results are interesting,
more importantly, this scenario of pressure and rotational support playing important roles in the steady-state solution more closely matches realistic disk scenarios that might arise from more complicated simulations. 
When pressure is eliminated entirely, these sorts of tests essentially only test the gravitational solver end of an algorithm, with some contribution from the hydrodynamics in that it should not distort the purely gravitational rotational flow.
As our examples here (and previous investigations of this Keplerian limit) demonstrate, many popular forms of SPH fail this limiting test due to the activation of the quadratic term of the artificial viscosity.
However, as we have also demonstrated, this is not the full story, and scenarios where both hydrodynamics and gravitational motion play non-negligible roles should be tested.

As stated previously, the momentum transport errors we observe in these tests are not trivial. 
At $t=200$, the $f_p=0.95$ disk has only experienced 7 orbits at its fastest annulus, and while this timescale is not the dynamically important one when pressure constitutes 95\% of the support against gravity, it nevertheless demonstrates that even just a few dynamical times are sufficient to entirely disrupt the balance between pressure and orbital motion.
This is best seen in \crefrange{fig:0p95v200t}{fig:0p50v200t} (for $f_p=0.95$ and $f_p=0.50$, respectively), where the M\&G simulations no longer resemble a $v^2\propto r^{-1}$ disk, and the Balsara and C\&D simulations are well on their way to the same fate.
Had we included secondary physics effects like radiation transport or chemical cooling, these errors would have resulted in the complete collapse of the disk. 
Since artificial viscosity converts momentum into thermal energy, and if thermal energy is rapidly being removed from the disk, along with wholesale momentum transport to the outer edges of the disk as must happen to maintain pressure balance, the disk would experience rapid orbital decay after only a few cooling timescales. 
Any real application that is designed to look for disk collapse, such as in galaxy formation simulations \citep[\eg][]{zurek1986,saitoh2008,inoue2014}, would therefore feature overly vigorous collapse arising purely from numerical errors, rather than actual, physical processes.

In the case of compact object merger disks, where our $f_p=0.50$ simulations have the greatest insight, these disks are assumed to be optically thick, and so cooling mechanisms are often excluded from the simulations. 
For those simulations, artificial momentum transport away from the compact object near the center of the disk still results in an unphysically short time-scale for disk collapse.
Many investigators performing simulations of this kind \citep{rosswog2010,vanKerkwijk2010,raskin2010,raskin2012,raskin2014,moll2014} have looked for conditions for carbon ignition during the post-merger, disk relaxation phases, and this effect may play an outsized role in the conclusions they reach.
Specifically, by observing the density curves of \cref{fig:0p50v200t}, we can conclude that any disk simulation employing some form of M\&G, Balsara, or C\&D viscosity will have experienced as much as a 50-75\% over-density near the center of the disk after only a handful of orbits. 
This has serious consequences for the expected lifetimes of these disks against carbon-ignition. 

Even for nearly pressure-free disks, as in our $f_p=0.05$ simulations, the small errors in the quadratic term of the artificial viscosity accumulate over time, and this effect may be difficult to examine in simulations where pressure plays no role at all.
In that case, the spurious heating from the artificial viscosity won't have destabilized the disk very much, and so such a disk will be more resilient against collapse.
Instead, the errors will accumulate to such a degree that the disk essentially falls apart, as has been witnessed by other investigators using negligible pressure \citep[\eg][]{cullen2010,hopkins2015,hosono2016}.

To conclude this investigation, we believe the generalized Keplerian disk including non-negligible pressure support is an important test problem that codes intended for use in modeling astrophysical problems involving gravitation and hydrodynamics should examine.  
Our specific results here -- testing how SPH and CRKSPH fare on these problems -- is interesting, but more broadly we would be interested in seeing many more astrophysical hydrodynamics methods tested against these scenarios, such as Eulerian meshed methods, moving mesh Lagrangian methods, etc.

\acknowledgments
This work was performed under the auspices of the U.S. Department of Energy by Lawrence Livermore National Laboratory under Contract DE-AC52-07NA27344.  
All of the calculations performed in this work leveraged \textsc{Spheral++}, an open-source SPH code freely available on SourceForge.
The source code includes the input scripts to reproduce the examples in this paper along with the solutions.
We would like to acknowledge the lyrical poeticism of Dead or Alive, whose 1985 hit played on repeat throughout the entirety of our investigation. 
\software{\textsc{Spheral++}, \url{https://sourceforge.net/projects/spheral}}

\bibliographystyle{apjsingle}

\end{document}